\documentclass[english,aps,twocolumn]{revtex4}
\usepackage[T1]{fontenc}
\usepackage[latin9]{inputenc}
\usepackage{textcomp}
\usepackage{amstext}
\usepackage{graphicx}
\usepackage{esint}

\makeatletter

\@ifundefined{textcolor}{}
{%
 \definecolor{BLACK}{gray}{0}
 \definecolor{WHITE}{gray}{1}
 \definecolor{RED}{rgb}{1,0,0}
 \definecolor{GREEN}{rgb}{0,1,0}
 \definecolor{BLUE}{rgb}{0,0,1}
 \definecolor{CYAN}{cmyk}{1,0,0,0}
 \definecolor{MAGENTA}{cmyk}{0,1,0,0}
 \definecolor{YELLOW}{cmyk}{0,0,1,0}
 }

\makeatother

\usepackage{babel}

\begin{document}

\preprint{This line only printed with preprint option}

\title{Separable approximation to two-body matrix elements}

\author{Luis M. Robledo}

\email{luis.robledo@uam.es}

\homepage{http://gamma.ft.uam.es/~robledo}

\affiliation{Departamento de F\'\i sica Te\'orica, Facultad de Ciencias M\'odulo 15, Universidad
Aut\'onoma de Madrid, E-28049 Madrid, Spain}

\date{\today}
\begin{abstract}
Two-body matrix elements of arbitrary local interactions
are written as the sum of separable terms in a way that is
well suited for the exchange and pairing channels present in
mean-field calculations.
The expansion relies on the transformation to center of mass and relative
coordinate (in the spirit of Talmi's method) and therefore it is only
useful (finite number of expansion terms) for harmonic oscillator
single particle states. The converge of the expansion with the number
of terms retained is studied for a Gaussian two body interaction.
The limit of a contact (delta) force is also considered. Ways to handle
the general case are also discussed.
\end{abstract}
\maketitle

\section{Introduction}

The evaluation of the pairing field in mean field theories like Hartree-
Fock plus Bardeen-Cooper-Schrieffer (HF+BCS) or Hartree-Fock- Bogoliubov
(HFB) is a computationally intensive task due to the non-local character
of the pairing tensor and the effort is comparable to the one devoted
to the evaluation of the exchange field in the Hartree Fock method.
Therefore, zero range pairing interactions are thoroughly used in
standard mean field calculations in order to reduce the computational
burden (see Ref \cite{Bender.03} for a recent review). The price
to pay for the use of zero range pairing forces is the introduction
of an {}``active space'' around the Fermi level to cut away the
ultraviolet divergences inherent to any contact interaction (the pairing
matrix elements are independent of the momentum transfer in nuclear
matter). Thus, it is customary to take into account in the gap equation
only those single particle levels lying inside a so called {}``active
window'' around the Fermi level and whose (sharp or soft) boundaries
are defined using reasonable (but arbitrary) assumptions. The boundaries
of the window as well as the pairing interaction strength are usually
fixed locally but they are usually not allowed to vary as a function
of other relevant degrees of freedom like the quadrupole deformation
of the nucleus. As a consequence, the impact of the rigid definition
of the {}``pairing active window'' in some observable magnitudes
can be relevant (see, for instance, \cite{Samyn.05} for a discussion
on the impact of the window's size on fission barriers). From this
perspective, the use of finite range pairing forces seems to be quite
unavoidable and this is a common argument to praise the use of Gogny
type interactions \cite{Decharge.80} in HFB or HF+BCS like mean field
calculations \cite{Ring.80,Bender.03}. It has to be mentioned that
a renormalization scheme for the phenomenological contact pairing
interactions has been proposed in Refs \cite{Bulgac.02A,Bulgac.02B}
to cure the {}``active window'' problem. Recently, proposals to
use as pairing interaction in finite nuclei a realistic finite range
two body bare interaction \cite{Duguet.04} or a low momentum evolved
version of it \cite{Duguet.04,Duguet.08,Lesinski.09} have been discussed
in the literature: the essence of the proposals is to introduce a
separable interaction in momentum representation \cite{Duguet.04}
that is used to fit the realistic interaction (or a low momentum evolved
version of it) by resorting to fitting protocols involving nucleon-nucleon
phase shifts or pairing gaps in nuclear matter. As a consequence of
the separable character in momentum space the force is non-local in
real space but in a way that simplifies the evaluation of the matrix
elements needed for finite nuclei calculations \cite{Duguet.08,Lesinski.09}.
Separable interactions were also used as a way to tackle the solution
of the Lippman-Schwinger equation for the $T$ operator with realistic
nucleon-nucleon potentials \cite{Brown.76}. Recently, a separable
form in momentum space of the pairing interaction of a Gaussian two
body force (as in the Gogny force) has been proposed \cite{Tian.09A,Tian.09B,Tian.09C}
as an alternative to more standard approaches based on zero range
contact pairing interactions. As the approximation relies on the same
ideas of Refs. \cite{Duguet.08,Lesinski.09} a non-local form of the
approximate interaction in real space is obtained. The non locality
of the interaction suggests the introduction of the Talmi-Brody-Moshinsky
transformation \cite{Talmi.59,Moshinsky.59,Brody.65} to center of
mass and relative coordinate to simplify the evaluation of the pairing
matrix elements in the harmonic oscillator basis. Explicit expression
for the matrix elements are given in \cite{Tian.09A} for an harmonic
oscillator basis with spherical quantum numbers and a Gaussian interaction.
Next, it has been shown that this new form is consistent with the
RPA framework \cite{Tian.09B}. Finally, the scheme has been extended
\cite{Tian.09C} to the case of a harmonic oscillator basis with axial
quantum numbers. The extension to the case of a harmonic oscillator
basis with triaxial quantum numbers is straightforward. In the present
article I show that the transformation to center of mass and relative
coordinates can be used for general two body interactions to obtain
a kind of separable expression for the pairing two body matrix elements
which involves the one body matrix elements of the interaction for
the relative coordinate wave functions. The use of the spectral representation
of this interaction matrix leads to a more explicit separable form
of the kind considered in \cite{Duguet.04,Duguet.08,Lesinski.09,Tian.09A,Tian.09B,Tian.09C}.
By using two specific examples in one dimension it is shown that the
number of separable terms in the expansion can be severely cut down
for short range interactions. The validity of the expansion concerning
ultraviolet divergences is also discussed. Finally, computational
schemes to deal with the Yukawa (and Coulomb!) potential are discussed.
It has to be stressed that the present approach is not limited to
pairing matrix elements and can  easily be extended to deal with the
exchange matrix elements required for the evaluation of the Fock potential
and therefore it could be used as an approximation scheme to shorten
the numerical burden of HFB calculations with finite range forces
like Gogny \cite{Decharge.80}, other recent proposals based on the
Yukawa potential \cite{Nakada.03} and even the Coulomb potential.

\section{Separable approximation to two-body matrix elements}

\subsection{General procedure}

Let us consider the (not antisymmetrized) two-body matrix element 
$\nu_{n_{1},n_{2},n_{3},n_{4}}=\int d\vec{r}_{1}\int d\vec{r}_{2}\phi_{n_{1}}^{*}
(\vec{r}_{1})\phi_{n_{2}}^{*}(\vec{r}_{2})v(\vec{r}_{1}-\vec{r}_{2})\phi_{n_{3}}
(\vec{r}_{1})\phi_{n_{4}}(\vec{r}_{2})$
for harmonic oscillator (HO) wave functions $\phi_{n}(\vec{r})$ (this
is not a fundamental restriction as more general single particle wave
functions can always be expanded in a HO basis). Taking into account
that at the end we will be folding this matrix elements with a matrix
(either the pairing tensor or the density for the exchange potential) 
with indexes $n_{3}$ and $n_{4}$ the best approach for the evaluation 
of this matrix element is that of Talmi \cite{Talmi.59} and Brody-Moshinsky 
\cite{Moshinsky.59,Brody.65} (see also \cite{deShalit.63,Irvine.72} for a 
general discussion).
Within this method, the product of HO single particle wave functions
with different arguments $\phi_{n_{3}}(\vec{r}_{1})\phi_{n_{4}}(\vec{r}_{2})$
is expanded in terms of suitable center of mass $\vec{R}=\frac{1}{\sqrt{2}}(\vec{r}_{1}+\vec{r}_{2})$
and relative coordinate $\vec{r}=\frac{1}{\sqrt{2}}(\vec{r}_{1}-\vec{r}_{2})$
wave functions (this unusual definition of the center of mass and
relative coordinate renders some of the expressions simpler than with
the usual definition)\begin{equation}
\phi_{n_{3}}(\vec{r}_{1})\phi_{n_{4}}(\vec{r}_{2})=\sum_{Nn}M_{n_{3}n_{4}}^{Nn}\phi_{N}(\vec{R})\phi_{n}(\vec{r})\label{eq:Mosh}\end{equation}
The center of mass and relative coordinate wave functions have the
same structural form as the original wave functions and the expansion
is finite (the range of values of $N$ and $n$ is finite !). This
is a direct consequence of the special structure of the HO
wave functions (the product of a Gaussian times a polynomial). Apart
from the HO, this peculiarity is only preserved in the case of plane
waves and the developments considered below can be  extended easily
to this case too. The expansion coefficients $M_{n_{1}n_{2}}^{Nn}$
are referred to as Talmi-Brody-Moshinsky \cite{Talmi.59,Moshinsky.59,Brody.65}
coefficients (TBMC). The TBMC have a selection rule that will help
to reduce the number of final separable terms, it is $n_{1}+n_{2}=N+n$
(see Appendix A for the general expression and selection rules in
the one-dimensional case) which implies that only one of the two sums
in Eq. (\ref{eq:MoshS}) is relevant. Explicit forms of the TBMC coefficients
have been discussed several times in the literature both in the 3D
spherical form of the HO wave functions \cite{Moshinsky.59,Talmi.59,Brody.65}
as in the one-dimensional case \cite{Talmi1D} . Introducing the expansion
of Eq. (\ref{eq:MoshS}) into the definition of the matrix element
we obtain the result\begin{equation}
\nu_{n_{1},n_{2},n_{3},n_{4}}=\sum_{N}\sum_{nn'}M_{n_{1}n_{2}}^{*\, Nn}v_{nn'}M_{n_{3}n_{4}}^{Nn'},\label{eq:me-1}\end{equation}
where we have made use of the orthogonality of the center of mass
wave functions $\phi_{N}(\vec{R})$ and introduced the matrix elements\begin{equation}
v_{nn'}=\int d\vec{r}\phi_{n}^{*}(\vec{r})v(\sqrt{2}\vec{r})\phi_{n'}(\vec{r}).\label{eq:vnm}\end{equation}
Please note the $\sqrt{2}$ factor in the argument of the interaction
that is a direct consequence of the definition of the relative coordinate.
This result can be easily generalized to non-local interactions of
the form\begin{equation}
\langle\vec{r}_{1}\vec{r}_{2}|\hat{v}|\vec{r}_{1}'\vec{r}_{2}'\rangle=\delta(\vec{R}-\vec{R}')v(\vec{r},\vec{r}')\label{eq:GenV}\end{equation}
where $\vec{R}$ and $\vec{r}$ are the center of mass and relative
coordinates associated to $\vec{r}_{1}'$ and $\vec{r}_{2}'$. Applying
the transformation of Eq. (\ref{eq:Mosh}) we obtain for the HO matrix
elements the same expression as in Eq. (\ref{eq:me-1}) but replacing
$v_{nn'}$ of Eq. (\ref{eq:vnm}) by
\begin{equation}
v_{nn'}=\int d\vec{r}\int d\vec{r}\,'\phi_{n}^{*}(\vec{r})v(\vec{r},\vec{r}\,')\phi_{n'}(\vec{r}\,').\label{eq:vnmG}
\end{equation}

The $v_{nn'}$ are matrix elements of a hermitian matrix and therefore
can be written, by resorting to the spectral decomposition of the
matrix, as \begin{equation}
v_{nn'}=\sum_{L}D_{nL}^{*}v_{L}D_{n'L}\label{eq:vnnD}\end{equation}
where the coefficients $v_{L}$ (the eigenvalues of $v_{nn'}$) are
real quantities. By introducing the coefficients\begin{equation}
\tilde{M}_{n_{1}n_{2}}^{NL}=\sum_{n}M_{n_{1}n_{2}}^{Nn}D_{nL}\label{eq:Mtilda}\end{equation}
we can finally cast Eq. (\ref{eq:me-1}) in a form that corresponds clearly
to a separable expansion for pairing and exchange matrix elements
as\begin{equation}
\nu_{n_{1},n_{2},n_{3},n_{4}}=\sum_{N}\sum_{L}\tilde{M}_{n_{1}n_{2}}^{*\, NL}v_{L}\tilde{M}_{n_{3}n_{4}}^{NL}.\label{eq:ME-Sep}\end{equation}
This separable expansion is still exact and can be used to express
the pairing field also as a sum of separable terms\[
\Delta_{n_{1}n_{2}}=\sum_{n_{3}n_{4}}\nu_{n_{1},n_{2},n_{3},n_{4}}\kappa_{n_{3}n_{4}}=\sum_{NL}\tilde{M}_{n_{1}n_{2}}^{*\, NL}v_{L}\Lambda_{NL}\]
with $\Lambda_{NL}=\sum_{n_{3}n_{4}}\tilde{M}_{n_{3}n_{4}}^{NL}\kappa_{n_{3}n_{4}}$.
At this point it is worth mentioning that this procedure can
be straightforwardly  extended to the evaluation of the exchange field
by introducing the new TBMC as \begin{equation}
\phi_{n_{1}}^{*}(\vec{r}_{1})\phi_{n_{4}}(\vec{r}_{2})=\sum_{Nn}\bar{M}_{n_{1}n_{4}}^{Nn}\phi_{N}(\vec{R})\phi_{n}(\vec{r})\label{eq:MoshS}\end{equation}
to obtain\[
\nu_{n_{1},n_{2},n_{3},n_{4}}=\sum_{N}\sum_{nn'}\bar{M}_{n_{1}n_{4}}^{Nn}v_{nn'}\bar{M}_{n_{2}n_{3}}^{*\, Nn'}\]
which can be used for the evaluation of the exchange field. This possibility
has not been considered in Refs. \cite{Tian.09A,Tian.09C}. 

The drawback of the (still exact) separable expansion being considered
is that the number of terms involved is enormous unless some property
of the interaction is invoked to restrict it. To do so, we can use
the general ideas of linear algebra on how to approximate an arbitrary
matrix in terms of low rank Kronecker tensor products (see \cite{VanLoad.92,VanLoan.00}
for an introduction to the subject) to justify the rank one approximation
\begin{equation}
v_{nn'}\approx d_{n}^{*}d_{n'}\label{eq:ZeroOrder}\end{equation}
where $d_{n}=\sqrt{v_{0}}D_{n0}$ and we are assuming that the $L=0$
term in Eq. (\ref{eq:vnnD}) corresponds to the largest eigenvalue
$v_{0}$ of the interaction matrix $v_{nn'}$. In this way we have
reduced the number of separable terms by a factor that corresponds
to the dimension of the matrix $v_{nn'}$. For better accuracy, we
can pursue further this idea to suggest the approximation\begin{equation}
v_{nn'}\approx\sum_{L=0}^{L_{C}}D_{nL}^{*}v_{L}D_{n'L}\label{eq:L0order}\end{equation}
(where $v_{L}$ with $L=0,\ldots,L_{C}$ stand for the $L_{C}+1$
largest eigenvalues of the matrix $v_{nn'}$). Obviously, low rank
Kronecker tensor products do not necessarily have to be based on the
eigenvectors of the interaction matrix $v$ and other alternatives
inspired in the physics to be described could be incorporated easily
(a possibility that comes up immediately is to use for $d_{n}$ in
Eq. (\ref{eq:ZeroOrder}) a linear combination of the eigenvectors
of $v$ with weights physically inspired by the problem to be treated
and to be fitted for the optimal reproduction of some quantity.)

\subsection{Zero range interaction}

In the following we will consider some typical interactions to asses
the validity of the preceding approximation scheme and we will start by
considering a zero range contact interaction $v(\vec{r})=G\delta(\vec{r})$.
In this case, the matrix $v_{nn'}$ is simply given by \begin{equation}
v_{nn'}=\frac{G}{2^{3/2}}\phi_{n}^{*}(0)\phi_{n'}(0)\label{eq:vnnDelta}\end{equation}
 which implies, in the language of Eq. (\ref{eq:vnnD}), that all
the eigenvalues $v_{L}$ are zero except the $L=0$ one. For this
interaction, the rank one approximation of Eq. (\ref{eq:ZeroOrder})
is exact and the reduction in the number of separable term is quite
significant. Introducing the quantity \[
\tilde{M}_{n_{1}n_{2}}^{N}=\sum_{n}M_{n_{1}n_{2}}^{Nn}\phi_{n}(0)\]
we finally obtain

\begin{equation}
\nu_{n_{1},n_{2},n_{3},n_{4}}=\frac{G}{2^{3/2}}\sum_{N}\tilde{M}_{n_{1}n_{2}}^{*\, N}\tilde{M}_{n_{3}n_{4}}^{N}.\label{eq:DeltaExp}\end{equation}
Apart from the analytical result, this formula is telling us that
for short range interactions the expansion of Eq. (\ref{eq:L0order})
is expected to be accurate enough by considering only a limited number
of terms.

In order to get a deeper insight into the number of separable terms
in the preceding expansion for the zero range force we will focus on the
one-dimensional case. Our basis will be the one-dimensional harmonic
oscillator basis $\phi_{n}(x)$ with $n=0,\ldots,M$ and containing
$M+1$ elements. The TBM expansion of the 1-D HO wave functions is
an expansion of the product of two polynomials in the variables $x_{1}$
and $x_{2}$ in terms of the product of polynomials in the variables
$X=\frac{1}{\sqrt{2}}(x_{1}+x_{2})$ and $x=\frac{1}{\sqrt{2}}(x_{1}-x_{2})$.
Although its explicit expression is known since a long time \cite{Talmi1D},
we present in Appendix A a brief derivation based on the generating
function of the Hermite polynomials. As a consequence of the explicit
form of the TBM coefficients, both quantum numbers $N$ (center of
mass) and $n$ (relative) should range from 0 up to $2M$ (i.e. $2M+1$
possible values) if none of the selection rules of the TBM coefficients
is taken into account. The parity selection rule $(-1)^{n_{1}+n_{2}}=(-1)^{N+n}$
of the TBM coefficients can be used to reduce the number of terms
in the $N$ sum. As $\phi_{n}(0)$ is only different from zero for
even values of $n$, the parity of $N$ has to be the one of $n_{1}+n_{2}$
and therefore the number of terms in the $N$ sum of Eq. (\ref{eq:DeltaExp})
gets reduced to half the initial value $(M+1$ to be more precise).

\subsection{Gaussian interaction}

In order to analyze in more detail the consequences of the proposed
expansion, we will study the example of a Gaussian interaction of
range $\mu$ in one dimension. The more realistic three-dimensional
case can be elucidated from the results obtained here as, in that
case, the matrix elements are the product of the one-dimensional ones
along each of the three spatial dimensions. In addition, the zero
range interaction results can be recovered in the limit $\mu\rightarrow0$
providing further insight into the number of terms to be retained
as well as potential problems with ultraviolet divergences. The one-dimensional
HO wave functions will be denoted as $\varphi_{n}(z;b)=\exp(-\frac{1}{2}\frac{z^{2}}{b^{2}})\tilde{\varphi}_{n}(z/b)$
where $\tilde{\varphi}(z/b)$ is the polynomial part of the HO wave
function given by the product of the Hermite polynomial of degree
$n$ times the normalization constant of the 1D HO wave function.
The quantity to evaluate is \begin{eqnarray*}
v_{nn'} & = & \int dz\,\varphi_{n}^{*}(z;b)\exp(-\frac{2z^{2}}{\mu^{2}})\varphi_{n'}(z;b)\\
 & = & \int dz\,\tilde{\varphi}_{n}^{*}(z/b)\exp(-(\frac{2}{\mu^{2}}+\frac{1}{b^{2}})z^{2})\tilde{\varphi}_{n'}(z/b).\end{eqnarray*}
This integral can be carried out in many different ways but for our
purposes it is better to use the transformation matrix $D(b'/b)$
connecting $\tilde{\varphi}_{n}(z/b)$ with $\tilde{\varphi}_{n}(z/b')$\[
\tilde{\varphi}_{n}(z/b)=\sum_{n'=0}^{n}D_{nn'}(b'/b)\tilde{\varphi}_{n'}(z/b'),\]
defined in Appendix B, to write\[
v_{nn'}=\sum_{L}D_{nL}^{*}(B/b)D_{n'L}(B/b)\]
with $B/b=\mu/\sqrt{\mu^{2}+2b^{2}}=\eta$. It is better to write
the preceding result as
\begin{equation}
v_{nn'}=\sum_{L}\tilde{D}_{nL}^{*}(\eta)\eta^{2L+1}\tilde{D}_{n'L}(\eta)\label{eq:vnnGauss}\end{equation}
to make explicit the dependence in the range of the Gaussian $\mu$.
Although this expression is not an explicit power expansion in $\eta$
due to the dependence of $\tilde{D}$ on this parameter, it is close
to be so because because the ratio between the $L+2$ term of the
sum and the $L$ one is given by $\eta^4(1+\mu^2/(2b^2))^2 (n-L)(n'-L)/((L+2)(L+1))$
that shows a dominant $\eta^4$ behavior. As $\eta$ typically goes as $\mu/b$
we can conclude that Eq (\ref{eq:vnnGauss}) is a kind of expansion on the 
range of the interaction $\mu$.
For small ranges $\mu\rightarrow0$ the parameter $\eta$ tends to
$\mu/(\sqrt{2}b)$ and therefore the most significant term in the
sum is $L=0$. The remaining terms decrease as $\mu^{2L}$ and justify
to cut the expansion of Eq. (\ref{eq:vnnGauss}) at some small $L_{C}$
\begin{equation}
v_{nn'}^{\textrm{App}}=\sum_{L=0}^{L_{C}}\tilde{D}_{nL}^{*}(\eta)\eta^{2L+1}\tilde{D}_{n'L}(\eta).\label{eq:vnnGaussApp}\end{equation}
Using the connection between a Gaussian
of vanishing range and the delta function \[
\delta(x_{1}-x_{2})=\lim_{\mu\rightarrow0}\frac{1}{\sqrt{\pi}}\frac{e^{-(x_{1}-x_{2})^{2}/\mu^{2}}}{\mu}\]
it is possible to obtain Eq. (\ref{eq:vnnDelta}) from Eq. (\ref{eq:vnnGauss})
by appropriately taking the $\mu\rightarrow0$ limit. As can be 
shown easily, the $\mu\rightarrow0$ limit of $\tilde{D}_{n0}(\eta)$ (the
only remaining term is $L=0)$ is given by $\pi^{1/4}b^{1/2}\varphi_{n}(0)$
as required. 

\begin{figure}
\includegraphics{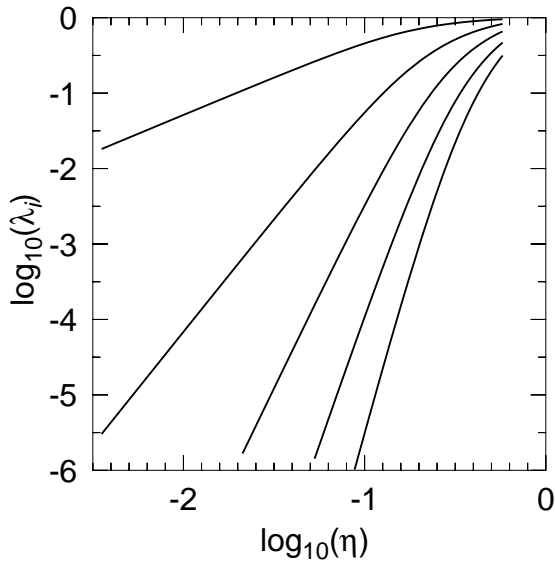}\caption{Largest eigenvalues $\lambda_{i}$ of the interaction matrix in the
case of a one-dimensional Gaussian interaction of range $\mu$. The
eigenvalues shown are plotted as a function of the parameter $\eta$
defined in the text. Notice the log-log character of the plot.\label{fig:lambdai}}

\end{figure}

At this point it is worth noticing that neither the eigenvalues nor
the eigenvectors of the $v_{nn'}$ matrix in the Gaussian case are
given by $\eta^{2L+1}$ or by the coefficients $\tilde{D}_{n'L}(\eta)$
above (see Eq. (\ref{eq:vnnGauss})). In order to understand the differences,
the largest eigenvalues obtained by numerical diagonalization in the
case $N=40$ and for $b=2$fm are plotted in Fig. \ref{fig:lambdai}
as a function of $\eta$. We observe in the log-log plot that for
small values of $\eta$ the eigenvalues have a linear behavior that
corresponds to $\eta^{2L+1}$ for $L=0,\ldots$ recovering in this
case the result of Eq. (\ref{eq:vnnGauss}). However, for larger values
of $\eta$ this is not the case indicating that the eigenvalues of
$v_{nn'}$ depart from the quantities in Eq. (\ref{eq:vnnGauss}).
An interesting observation regarding the behavior of the eigenvalues
is that they are a decreasing function as $\mu$ decreases, justifying
the idea that keeping only the largest eigenvalues is a kind of short
range expansion of the matrix elements. Obviously, if all the terms
are considered, both the expansion of Eq. (\ref{eq:vnnGauss}) and
the general one of the spectral decomposition of the matrix $v_{nn'}$
give the exact answer. The differences will show up in the approximate
case when only a finite number of terms is retained. We have checked
that the spectral decomposition always gives the best results if quantified
in terms of the mean square deviation $\sigma_{v}=\sqrt{\sum_{nn'}(v_{nn'}-v_{nn'}^{\textrm{App}})^{2}}/(N+1)$,
but none of the approximate matrix elements $v_{nn'}^{\textrm{App}}$
reproduce the exact ones, as it may happen when using Eq. (\ref{eq:vnnGaussApp}).
As a consequence of the selection rules of the $D$ coefficients involved,
the exact values of $v_{nn'}$ for low values of $n$ and $n'$ can
be obtained from Eq. (\ref{eq:vnnGaussApp}) even for small values
of $L_{C}$. This is not in contradiction with what it was said about
$\sigma_{v}$, as the approximation based on Eq. (\ref{eq:vnnGaussApp})
provides, for large values of $n$ and $n'$, worse matrix elements
than the ones obtained using the spectral decomposition of $v_{nn'}$.
The analysis below of the two body matrix elements has been carried
out in parallel using both possibilities and it has been found that,
although the spectral decomposition provides again a better overall
approximation for the two body matrix elements, those with small quantum
numbers (the most relevant for the physics of the system to be described)
are slightly better reproduced by the approximations based in Eq.
(\ref{eq:vnnGaussApp}). For the above reasons we will only show results
based on Eq. (\ref{eq:vnnGaussApp}). 

\begin{figure}
\includegraphics[width=0.95\columnwidth]{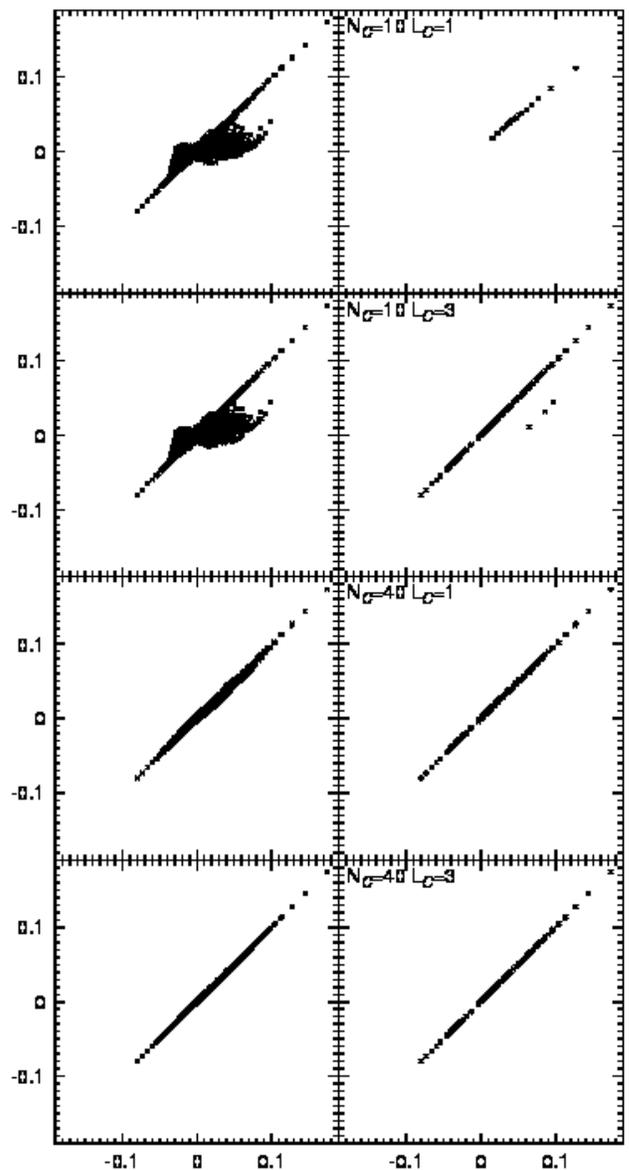}
\caption{Approximate matrix elements ($y$ axis) for different values of $L_{C}$
and $N_{C}$ are plotted versus the exact values ($x$ axis) for the
one-dimensional Gaussian interaction with range $\mu=0.7$fm and harmonic
oscillator length $b_{z}=$2.0 fm. On the left hand side panels, all
matrix elements with $n_{1},$ $n_{2},$ $n_{3}$ and $n_{4}$ ranging
from $0$ up to 20 (i.e. a total of $21^{4}$ matrix elements) are
depicted. On the right hand side panels only those matrix elements
with $n_{1},$ $n_{2},$ $n_{3}$ and $n_{4}$ smaller than 7 are
depicted (see text for details).\label{fig:Fig1}}

\end{figure}

\begin{figure}
\includegraphics[width=0.95\columnwidth]{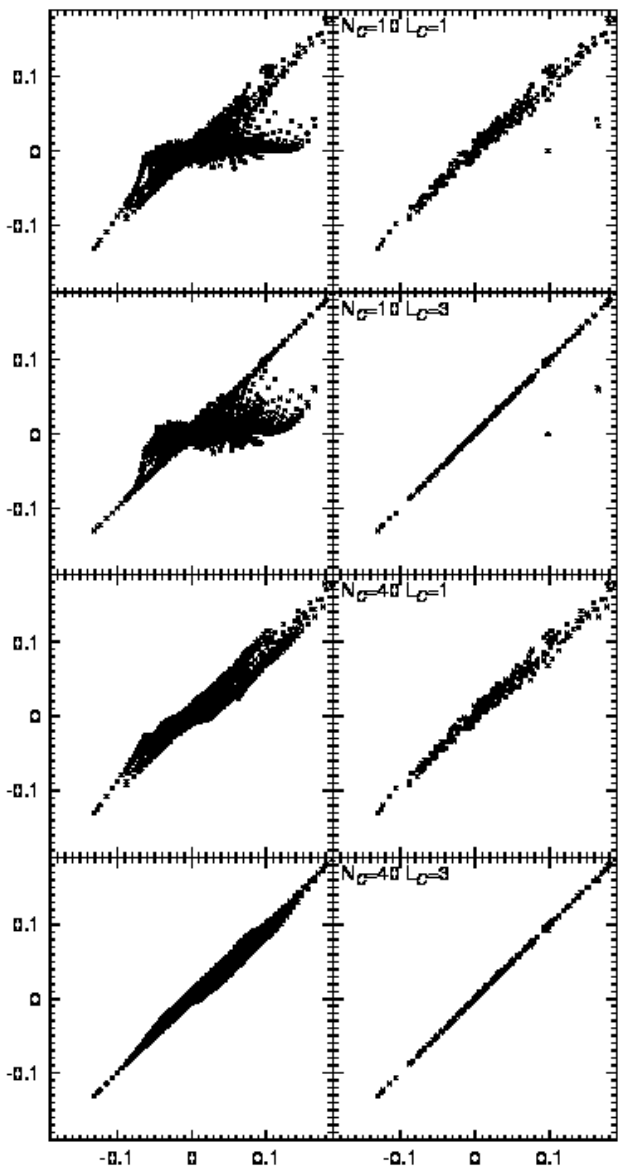}
\caption{Same as Fig. \ref{fig:Fig1} but for a range $\mu=1.4$fm of the Gaussian
interaction.\label{fig:Fig2}}

\end{figure}

The result obtained in Eq. (\ref{eq:vnnGaussApp}) for short ranges
$\mu$ suggests to approximate $v_{nn'}$ by keeping only the $L=0$
term. However, taking into account that the transformation coefficients
$D(\eta)$ preserve parity, we realize that the $L=0$ approximation
will make automatically zero the $v_{nn'}$ matrix elements with odd
$n$. Therefore, we have to include both $L=0$ and $L=1$ as the
leading order. The next to leading order will correspond to take $L=2$
and $3$ and so on. In order to test the convergence rate of such
an approximation we have performed calculations comparing the exact
matrix elements of the one-dimensional Gaussian interaction with those
obtained using Eq. (\ref{eq:ME-Sep}) and restricting the sum in $L$
from 0 up to a parameter denoted $L_{C}$ as well as the sum in $N$
from 0 up to $N_{C}$. The size of the one-dimensional HO basis considered
is of 21 shells ($n=0,\ldots,20$) and there are 97241 non zero matrix
elements. The results for a Gaussian of range $\mu=0.7$ fm are depicted
in Fig. \ref{fig:Fig1}. On the left hand side panels all the matrix
elements are plotted ($y$ axis, approximate values, $x$ axis exact
ones) for different values of $N_{C}$ and $L_{C}$. On the right
hand side panels, the same kind of plot is depicted but this time
only the matrix elements with $n_{1},$ $n_{2},$ $n_{3}$ and $n_{4}$
smaller than 7 are shown. The reasons are two: first that the realistic
single particle orbitals are usually close to the HO wave functions 
with the same quantum number $n$ and therefore and expansion on the HO
basis usually requires the $n-2$, $n$ and $n+2$ states for an reasonable
representation. The second reason is that
only a limited number of orbitals around the Fermi level usually play
a role in the pairing properties and therefore only the quantum numbers
$n$ required to accommodate $A$ particles are required. Usually seven
major shells (the corresponding quantity in the three-dimensional case)
are required to well accommodate of the order of 200 particles. 
Therefore, in the left hand side
panels there are many matrix elements that will contribute little
to the pairing field and can be safely disregarded for the discussion
of the quality of the approximation. That is, the matrix elements
depicted on the right hand side panels are supposed to be the most
relevant for the pairing properties of the nucleus. By looking at
the plot we observe how the bigger the values of $N_{C}$ and $L_{C}$
are the better the approximation is (the points gather around the
$y=x$ line). We even observe how the $N_{c}=10$ and $L_{C}=1$ approximation,
including eleven separable terms, is already quite reasonable for
the reduced set of relevant matrix elements (right hand side panels).
In Fig. \ref{fig:Fig2} the same kind of plots as the ones of Fig.
\ref{fig:Fig1} are presented but this time for a Gaussian twice the
previous range, i.e. $\mu=1.4$ fm. We immediately realize the worsening
of the approximation, what is consistent with the previous findings
that this is a kind of {}``short range'' expansion. In this Figure
we observe that the convergence with $N_{c}$ is much faster than
the one with $L_{C}$ because already the $N_{C}=10$ numbers compare quantitatively
well with the ones of $N_{C}=40$ whereas for $L_{C}$ the results
with $L_{C}=3$ are much better than the ones for $L_{C}=1$.

\begin{figure}
\includegraphics[width=0.95\columnwidth]{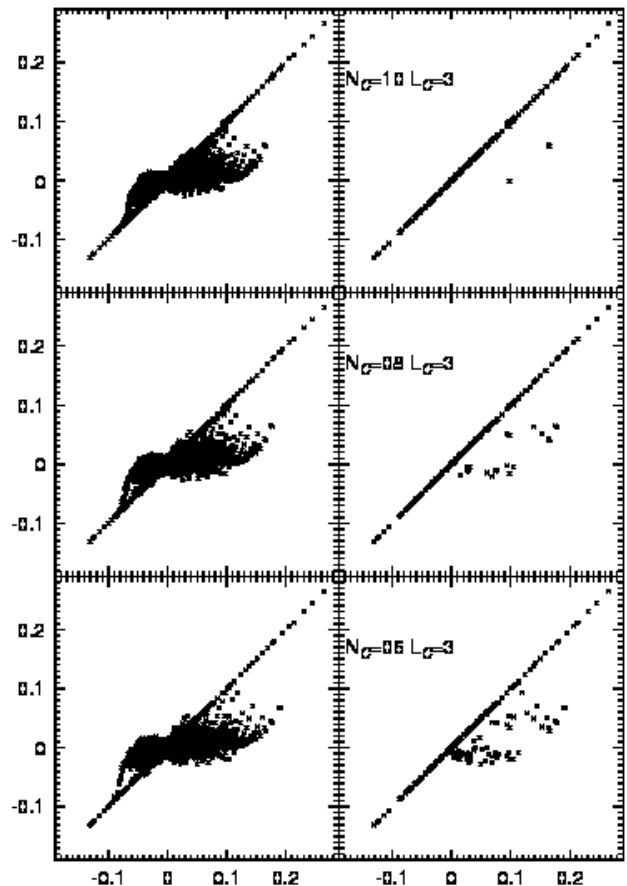}
\caption{Comparison between exact ($x$ axis) and approximate ($y$ axis) pairing
matrix elements of an one-dimensional Gaussian interaction with range
$\mu=1.4$ fm as a function of $N_{C}$ (see text for details as well
as Fig. \ref{fig:Fig1} for the meaning of the left hand side and
right hand side panels). \label{fig:Fig3}}

\end{figure}

Finally, in Fig. \ref{fig:Fig3} we present a more detailed study
of the convergence with $N_{C}$ in the case of $\mu=1.4$ fm keeping
fixed $L_{C}$ to the reasonable value of 3. We observe how decreasing
the value of $N_{C}$ from 10 to 8 to 6 degrades the quality of the
approximation for the relevant matrix elements (right hand side panels)
but this is not significant and it is quite likely that even $N_{C}=6$
(twelve separable terms!) will provide already reasonable values of
the pairing tensor in calculations with real nuclei. This point can
only be tested in realistic calculations and work along this direction
is in progress.

It is also interesting to compare the approach of Refs. \cite{Tian.09A,Tian.09B,Tian.09C,Duguet.04,Duguet.08,Lesinski.09}
with the present one. In those references the two body Gaussian interaction
is replaced by\begin{equation}
\langle\vec{r}_{1}\vec{r}_{2}|v|\vec{r}_{1}',\vec{r}_{2}'\rangle=G\delta(\vec{R}-\vec{R}\,')\sum_{\alpha}\lambda_{\alpha}P_{\alpha}(r)P_{\alpha}(r')\label{eq:SepNonLoc}\end{equation}
where $\vec{R}$ and $\vec{r}$ are the center of mass and relative
coordinate of $\vec{r}_{\text{1}}$ and $\vec{r}_{2}$. The functions
$P_{\alpha}(r)$ are taken as Gaussian with widths adjusted to reproduce
nucleon-nucleon phase shifts \cite{Duguet.04,Lesinski.09} or the
nuclear matter pairing gap \cite{Tian.09A,Tian.09B,Tian.09C}. More
involved expressions and fitting strategies were used in \cite{Duguet.08}.
This form of the interaction corresponds to the class introduced in
Eq. (\ref{eq:GenV}) and therefore we can use Eq. (\ref{eq:vnmG})
to obtain $v_{nn'}=\sum_{\alpha}\lambda_{\alpha}\left(P_{\alpha}\right)_{n}^{*}\left(P_{\alpha}\right)_{n'}$
where $\left(P_{\alpha}\right)_{n}=\int d\vec{r}\phi_{n}(\vec{r})P_{\alpha}(r)$.
For each of the $P_{\alpha}(r)$ factors, the corresponding term is
nothing but a rank one Kronecker approximation to the exact $v_{nn'}$
matrix of the interaction. The vectors $\left(P_{\alpha}\right)_{n}$
considered in Refs. \cite{Tian.09A,Tian.09B,Tian.09C,Duguet.04,Duguet.08,Lesinski.09}
are inspired by physical requirements and the free parameters entering
their definitions can be used (as it is the case) to find an optimal
(in physical terms) approximation to $v_{nn'}$ by the requirements
of reproducing as well as possible some nuclear properties. From our
previous numerical study we can conclude that our {}``blind'' approximation
is also well suited to deal with the problem.

\subsection{Ultraviolet divergence}

\begin{figure}

\includegraphics{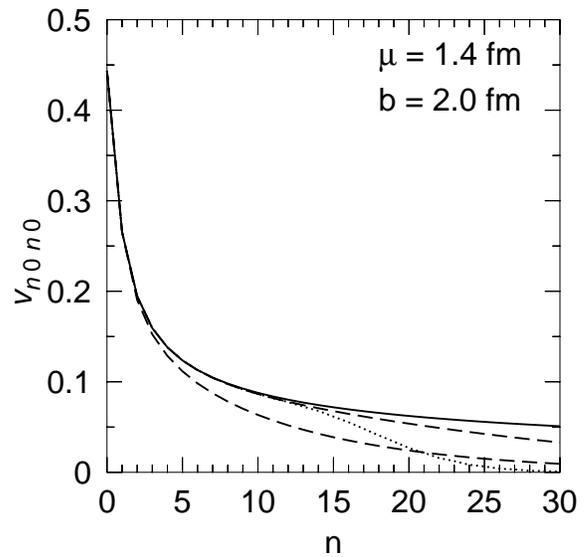}\caption{Matrix element $v_{n,0,n,0}$ for the one-dimensional 
Gaussian interaction
of range $\mu=1.4$ fm and oscillator length parameter $b=$2.0 fm
plotted as a function of $n$. Full curve corresponds to the exact
result, dashed ones to the approximation with $N_{C}=60$ and $L_{C}=1$
and $L_{C}=3$ (the one closer to the exact result). The dotted curve
corresponds to the $N_{C}=10$ and $L_{C}=3$ approximation.\label{fig:vn0n0} }

\end{figure}

From the previous results one could conclude that, because the low rank
separable expansion is a kind of {}``short range'' expansion, it
should show some sort of ultraviolet divergence like the pairing matrix
elements of a zero range contact interaction. In this section I will
show that this is not the case. First of all, I will remind the reader
about the origin of the ultraviolet divergence. It is a direct consequence
of the fact that the nuclear matter pairing matrix elements \begin{equation}
-G\langle\vec{k}_{1},\vec{k}_{2}|\delta(\vec{r}_{1}-\vec{r}_{1})|\vec{k}'_{1},\vec{k}'_{2}\rangle=-\frac{G}{(2\pi)^{3}2^{3/2}}\delta(\vec{K}-\vec{K}')\label{eq:NMPME}\end{equation}
of a contact interaction do not depend on the momentum transfer vector
$\vec{q}=\vec{k}-\vec{k}'$ (we have introduced the vectors $\vec{K}=\frac{1}{\sqrt{2}}(\vec{k}_{1}+\vec{k}_{2})$
and $\vec{k}=\frac{1}{\sqrt{2}}(\vec{k}_{1}-\vec{k}_{2})$) favoring
infinitely high momentum transfers unless some cut off is introduced.
In the one-dimensional HO case, the structure of
the general matrix elements is more difficult to visualize, but for
the sake of discussion, we can restrict to the specific matrix elements
$-G\langle n,0|\delta(x_{1}-x_{2})|n,0\rangle$ which, as can be 
shown easily using the preceding formulas, are independent of $n$ and given
by $-G/\sqrt{2}$. The constant value favors, as in the nuclear matter
case, the scattering of the $n=0$ state into high $n$ states. For
an one-dimensional Gaussian of range $\mu$ the expression of the
previous matrix element is not as simple as in the case of the zero
range force, and therefore it is preferable to plot it as a function
of $n$ in order to discuss its properties. In Fig. \ref{fig:vn0n0}
we have plotted the matrix elements $\langle n,0|\exp(-(x_{1}-x_{2})^{2}/\mu^{2})|n,0\rangle$
as a function of $n$ as well as three of the low rank approximations
corresponding to $N_{C}=60$ and $L_{C}=1$ and 3 and $N_{C}=10$
and $L_{C}=3$ (see caption for details). The conclusions from this
plot are that the matrix element is a decreasing function of $n$
quenching the promotion to high $n$ states and therefore the ultraviolet
divergence. Also, the similarities between the exact matrix elements
and the two typical low rank approximations with $N_{C}=60$ for all
values of $n$ are remarkable. In the case $N_{C}=10$ and $L_{C}=3$
the approximation is reasonable up to $n=15$ and from there on the
matrix elements are too small as compared to the exact ones. In any
case, we conclude that the low rank approximations maintain the characteristic
decreasing with $n$ of the matrix elements that is needed to prevent
the ultraviolet divergence. 

\subsection{General interactions}

As discussed previously the case of a general interaction involves
the evaluation of the matrix $v_{nn'}$ and its diagonalization. The
evaluation of $v_{nn'}$ can be carried out numerically using, for
instance, Gauss-Hermite integration thanks to the presence of the
HO wave functions in the matrix elements. The resulting formula is
also in the form of Eq. (\ref{eq:vnnD}) as can be inferred by using
an one-dimensional example\begin{eqnarray}
v_{nn'} & = & \int_{-\infty}^{\infty}dz\,\varphi_{n}^{*}(z;b)v(\sqrt{2}z)\varphi_{n'}(z;b)\nonumber \\
 & \approx & \sum_{i=-N_{H}}^{N_{H}}w_{i}\tilde{\varphi}_{n}^{*}(z_{i})v(\sqrt{2}b_{i}z_{i})\tilde{\varphi}_{n'}(z_{i})\nonumber \\
 & = & \sum_{i=-N_{H}}^{N_{H}}D_{ni}^{*}(w_{i}v(\sqrt{2}b_{i}z_{i}))D_{n'i}\label{eq:GaussInt}\end{eqnarray}
where $z_{i}$ and $w_{i}$ are the nodes and weights of the Gauss-Hermite
integration, $\tilde{\varphi}_{n'}(z)$ is the reduced HO wave function
defined above and the coefficients $D_{ni}=\tilde{\varphi}_{n}(z_{i})$.
Also in this case, if $v(r)$ is short range, the quantity $w_{i}v(\sqrt{2}b_{i}z_{i})$
will decrease as $z_{i}$ increases. In the case of a Gaussian interaction
this procedure will reproduce the exact answer and will provide an
alternative (but equivalent) description of the previous results for
the Gaussian. Another alternative to this procedure, that lends more
analytical results, is to represent the interaction to be treated
by means of its Gauss transform. The Gauss transform is a derivative
of the more popular Laplace transform, where a change of variables
allows to express given functions as linear combinations of Gaussian
with different widths. The best known example of this treatment is
that of the Yukawa potential \cite{ExpSum} \begin{equation}
\frac{e^{-\mu r}}{r}=\frac{2}{\sqrt{\pi}}\int_{0}^{\infty}e^{-r^{2}t^{2}-\mu\text{\texttwosuperior}/(4t^{2})}dt\label{eq:Yukawa}\end{equation}
that can straightforwardly be used to deal with the Coulomb interaction
by taking the $\mu\rightarrow0$ limit (see Ref \cite{Girod.83} for
an early application in nuclear physics). The advantages of this method
are first that the separable expansion for the Gaussian is known analytically
and can be used straight ahead and second that the Gaussian interaction
is separable along the three spatial directions allowing to treat
the problem as three uncorrelated one-dimensional problems one for
each spatial direction. The latter advantage is specially helpful
to deal with the Coulomb interaction. Given the long range of
the Coulomb force the number of terms required for an accurate
separable expansion of a general matrix element is expected to be 
larger than for a gaussian. However, the relevant Coulomb matrix
elements for the nuclear case involve single particle wave
functions which are located inside the nucleus and therefore
explore the Coulomb potential in the interior of the nucleus.
For those matrix elements, the Coulomb force can be considered
as a short range interaction with a range of the order of the size of the
nucleus and therefore the number of separable terms required
for an accurate representation are expected to be much smaller
than the ones required for a general matrix element of the Coulomb
interaction. 

\subsection{Density dependent forces}

In the applications of the HFB method it is common to find
terms in the interaction/energy functional that are 
referred to as "Density Dependent" (DD) terms and are given
by the general expression
\begin{equation}
v(\vec{r}_1,\vec{r}_2) = f(\vec{r})G(\vec{R})
\end{equation}
Usually, $f(\vec{r})=\delta(\sqrt{2}\vec{r})$ and
$G=\rho^\alpha$, that is, the density raised to some
(usually non integer) power $\alpha$. Applying the ideas
developed in this article it is easy to obtain in this case
\begin{equation}
\nu_{n_{1},n_{2},n_{3},n_{4}}=\sum_{K}\sum_{L}\tilde{M}_{n_{1}n_{2}}^{*\,
 KL}f_{L}g_K\tilde{M}_{n_{3}n_{4}}^{KL}.\label{eq:DD-Sep}\end{equation}
where we have introduced the coefficients
\begin{equation}
\tilde{M}_{n_{1}n_{2}}^{KL}=\sum_{Nn}M_{n_{1}n_{2}}^{Nn}D_{nL}E_{NK}\label{eq:MtildaDD}
\end{equation}
given in terms of the spectral decomposition of the matrices $f$ and $G$ with 
eigenvalues $f_L$ and $g_K$, respectively. In the common situation where
$f(\vec{r})=\delta(\sqrt{2}\vec{r})$ the $f_L$ are constant and equal to $1/2^{3/2}$
and $D_{nL}=\phi_n(0)$ independent of $L$.

\section{Conclusions}

By means of the transformation properties of the HO
basis to the center of mass and relative coordinate a separable expansion
of the pairing and exchange matrix elements of a general two body
interaction is obtained. The study of two specific examples: the contact
delta force and the one-dimensional Gaussian interaction show that
the number of terms of the separable expansion to be considered can
be substantially reduced without affecting too much the accuracy of
the approximate formula. The separable expansion turns out to be a
kind of {}``short range'' expansion. The issue of ultraviolet divergences
inherent to any short range expansion is analyzed. The proposed separable
expansion opens up a new avenue to deal with finite range interactions
both in the pairing and exchange channel. In the former case, the
cumbersome definition of the {}``cut off window'' is avoided by
the implicit finite range. The possibility to extend these considerations
to other finite range interactions like the Coulomb potential is also
discussed.
\begin{acknowledgments}
Work supported in part by MEC (FPA2007-66069) and by the Consolider-Ingenio
2010 program CPAN (CSD2007-00042). Helpful suggestions and comments
by Th. Duguet and Th. Lesinski are gratefully acknowledged.
\end{acknowledgments}
\appendix

\section{Talmi-Brody-Moshinsky coefficients in one dimension}

The Talmi-Brody-Moshinsky coefficients in one dimension are defined
as \[
\varphi_{n_{1}}(x_{1})\varphi_{n_{2}}(x_{2})=\sum_{Nn}M_{n_{1}n_{2}}^{Nn}\varphi_{N}(X)\varphi_{n}(x)\]
where $X=\frac{1}{\sqrt{2}}(x_{1}+x_{2})$ and $x=\frac{1}{\sqrt{2}}(x_{1}-x_{2})$
are the center of mass and relative coordinate respectively and $\varphi_{n}(x)=(\sqrt{\pi}2^{n}n!b)^{-1/2}H_{n}(x/b)\exp(-\frac{1}{2}\frac{x^{2}}{b^{2}})$
is the one-dimensional HO wave function written in terms of the Hermite
polynomials. Explicit expression of $M_{n_{1}n_{2}}^{Nn}$ have already
being obtained in the literature \cite{Talmi1D} but we will introduce
here another derivation which is simple and straightforward. We will
take advantage of the generating function of the one-dimensional HO
wave functions \[
G(x/b,t)=\exp(-\frac{1}{2}\frac{x^{2}}{b^{2}}+2\frac{x}{b}t-t^{2})=\sum_{n}\chi_{n}(t)\varphi_{n}(x)\]
with $\chi_{n}(t)=(\sqrt{\pi}b2^{n}n!)^{1/2}t^{n}/n!$. Given the
Gaussian form of the generating function we have $G(x_{1}/b,t_{1})G(x_{2}/b,t_{2})=G(X/b,T)G(x/b,t)$
where $T=\frac{1}{\sqrt{2}}(t_{1}+t_{2})$ and $t=\frac{1}{\sqrt{2}}(t_{1}-t_{2})$.
Expanding in right hand side of the last identity $\chi_{N}(T)$ and
$\chi_{n}(t)$ in powers of $t_{1}$and $t_{2}$and comparing with
the same powers in the left hand side we finally obtain\begin{eqnarray*}
M_{n_{1}n_{2}}^{Nn} & = & \delta_{n_{1}+n_{2},N+n}\left(\frac{n_{1}!n_{2}!}{2^{n_{1}+n_{2}}N!n!}\right)^{1/2}\\
 & \times & \sum_{m}(-1)^{m}\left(\begin{array}{c}
N\\
n_{1}-n+m\end{array}\right)\left(\begin{array}{c}
n\\
m\end{array}\right)\end{eqnarray*}
where the selection rule is made explicit.

\section{Transformation coefficients}

In this appendix I supply the expression for the expansion coefficients
of the one-dimensional reduced HO wave function $\tilde{\varphi}_{n}(x/b)=\exp(\frac{1}{2}\frac{x^{2}}{b^{2}})\varphi_{n}(x/b)=(\sqrt{\pi}2^{n}n!b)^{-1/2}H_{n}(x/b)$
in terms of the same object but for a different length scale $b'$\begin{equation}
\tilde{\varphi}_{n}(x/b)=\sum_{n'=0}^{n}D_{nn'}\left(\frac{b'}{b}\right)\tilde{\varphi}_{n'}(x/b').\label{eq:Ddef}\end{equation}
There are many ways to obtain the $D$ coefficients although probably
the more economical one is to use the generating function of the Hermite
polynomials $H_{n}(x/b)$ which is given by (see previous appendix)
\begin{equation}
\tilde{G}(x/b,t)=\exp\left(2\frac{x}{b}t-t^{2}\right)=\sum_{n}\frac{t^{n}}{n!}H_{n}(x/b)\label{eq:GenF}\end{equation}
Using the generating function we can write $\tilde{G}(x/b,t)=\tilde{G}(x/b',t')\exp(-(1-\eta^{2})t^{2})$
where $\eta=b'/b$ and $t'=\eta t$. Equating equal powers in $t$
in the previous expression and using Eq. (\ref{eq:GenF}) we obtain\[
H_{n}(x/b)=\sum_{n'}(-1)^{\frac{n-n'}{2}}\frac{n!\eta^{n'}(1-\eta^{2})^{\frac{n-n'}{2}}}{n'!\left(\frac{n-n'}{2}\right)!}\Delta_{n,n'}H_{n'}(x/b')\]
which is an identity that finally yields to\begin{equation}
D_{nn'}(\eta)=\Delta_{n,n'}(-1)^{\frac{n-n'}{2}}\left(\frac{n!}{n'!}\right)^{1/2}\frac{\eta^{n'+1/2}(1-\eta^{2})^{\frac{n-n'}{2}}}{2^{\frac{n-n'}{2}}\left(\frac{n-n'}{2}\right)!}.\label{eq:Deta}\end{equation}
In this expression, the function $\Delta_{n,n'}=\frac{1}{2}(1+(-1)^{n+n'})$
which is one if $n$ and $n'$ have the same parity (even or odd)
and zero otherwise has been introduced. Notice also that $D_{nn'}(\eta)=0$
if $n'>n$ in agreement with the limits of the sum in Eq. (\ref{eq:Ddef}).


\begin{thebibliography}{24}
\bibitem{Bender.03}M. Bender, P.-H. Heenen and P.-G. Reinhard, \textit{{}``Self-consistent
mean-field models for nuclear structure''}, Rev. Mod. Phys. \textbf{75},
121 (2003) 

\bibitem{Samyn.05}M. Samyn, S. Goriely and J.M. Pearson, \textit{{}``Further
explorations of Skyrme-Hartree-Fock-Bogoliubov mass formulas. V. Extension
to fission barriers''}, Phys. Rev. C\textbf{72}, 044316 (2005)

\bibitem{Decharge.80}J. Decharg\'e and D. Gogny, \textit{{}``Hartree-Fock-Bogolyubov
calculations with the D1 effective interaction on spherical nuclei''},
Phys. Rev. C\textbf{21}, 1568 (1980)

\bibitem{Ring.80}P. Ring and P. Shuck, \emph{The Nuclear Many Body
Problem} (Springer--Verlag Edt. Berlin, 1980). 

\bibitem{Bulgac.02A}Aurel Bulgac and Yongle Yu, \textit{{}``Renormalization
of the Hartree- Fock- Bogoliubov Equations in the case of a zero range
pairing interaction''}, Phys. Rev. Lett. \textbf{88}, 042504 (2002)

\bibitem{Bulgac.02B}Aurel Bulgac, \textit{{}``Local density approximation
for systems with pairing correlations''}, Phys. Rev. C\textbf{65},
051305(R) (2002)

\bibitem{Duguet.04}T. Duguet, \textit{{}``Bare vs effective pairing
forces: A microscopic finite-range interaction for Hartree-Fock-Bogolyubov
calculations in coordinate space}'', Phys. Rev. C\textbf{69}, 054317
(2004)

\bibitem{Duguet.08}T. Duguet and T. Lesinski, {}``\textit{Non-empirical
pairing functional}'', Eur. Phys. J. ST \textbf{156}, 207 (2008)

\bibitem{Lesinski.09}T. Lesinski, T. Duguet, K. Bennaceur and J.
Meyer {}``\textit{Non-empirical pairing energy density functional''},
Eur. Phys. J. A\textbf{40,} 121 (2009)

\bibitem{Brown.76}G. E. Brown and A.D. Jackson, \textit{The nucleon-nucleon
interaction}, Chapter V, North-Holland (Amsterdam, 1976)

\bibitem{Tian.09A}Y. Tian, Z.Y. Ma and P. Ring, \textit{{}``A finite
range pairing force for density functional theory in superfluid nuclei''},
Phys. Lett. B\textbf{676}, 44 (2009)

\bibitem{Tian.09B}Yuan Tian, Zhong-yu Ma and Peter Ring \textit{{}``Separable
pairing force for relativistic quasiparticle random phase approximation''},
Phys. Rev. C\textbf{79}, 064301 (2009)

\bibitem{Tian.09C}Yuan Tian, Zhong-yu Ma and Peter Ring \textit{{}``Axially
deformed relativistic Hartree Bogoliubov with separable pairing force''},
Phys. Rev. C\textbf{80}, 024313 (2009)

\bibitem{Talmi.59}I. Talmi, \textit{{}``Nuclear spectroscopy with
harmonic-oscillator wave-functions''}, Helv. Phys. Acta, \textbf{25},
185 (1952)

\bibitem{Moshinsky.59}M. Moshinsky, \textit{{}``Transformation brackets
for harmonic oscillator functions''}, Nucl. Phys. \textbf{13}, 104
(1959)

\bibitem{Brody.65}T.A. Brody and M. Moshinsky, \textit{Tables of
Transformation Brackets for Nuclear Shell-Model calculations} (Universidad
Nacional de Mxico, Mxico, 1965)

\bibitem{Nakada.03}H. Nakada, \textit{{}``Hartree-Fock calculations
on unstable nuclei with several types of effective interactions''},
Nucl. Phys. \textbf{A722}, 117 (2003)

\bibitem{deShalit.63}A. de Sahlit and I. Talmi, \textit{Nuclear Shell
Theory}, Academic Press (New York, 1963)

\bibitem{Irvine.72}J.M. Irvine, \textit{Nuclear Structure theory},
Pergamon Press (Oxford, 1972) 

\bibitem{Talmi1D}Yu. F. Smirnov, {}``\textit{Talmi transformation
for particles with different masses}'', Nucl. Phys. \textbf{39},
346 (1962); L.A. Copley and A.B. Volkov, \textit{{}``A general two-body
matrix element for a Gaussian interaction using cylindrical harmonic
oscillator functions}'', Nucl. Phys. \textbf{84}, 417 (1966); R.R.
Chasman and S. Wahlborn, \textit{{}``Transformation scheme for harmonic-oscillator
wave functions}'', Nucl. Phys. \textbf{A90}, 401 (1967); R. Muthukrishnan,
\textit{{}``Intrinsic states of deformed nuclei in the Hartree-Fock
(HF) approximation''}, Nucl. Phys. \textbf{A93}, 417 (1967)

\bibitem{VanLoad.92}C.F. Van Loan and N.P. Pitsianis, {}``Approximation
with Kronecker products'', Linear Algebra for Large Scale and Real
Time Applications, M.S. Moonen, G.H. Golub (Eds.), p 293 (Kluwer Publications,
Dordrecht 1992) 

\bibitem{VanLoan.00}C.F. Van Loan, {}``\textit{The ubiquitous Kronecker
product}'', Jour. of Comp. and Appl. Math. \textbf{123}, 85 (2000)

\bibitem{ExpSum}R.J. Harrison, G.I. Fann, K. Yanai and G. Beylkin,
\textit{{}``Multiresolution Quantum Chemistry in Multiwavelet Bases''},
Lecture Notes on Computer Science \textbf{2660}, 103 (2003); G. Beylkin
and L. Monzn, \textit{{}``On approximation of functions by exponential
sums}'', Appl. Comput. Harmon. Anal. 19, 17 (2005)

\bibitem{Girod.83}M. Girod and B. Grammaticos, {}``\textit{Triaxial
Hartree-Fock-Bogolyubov calculations with D1 effective interaction}'',
Phys. Rev. C \textbf{27}, 2317 (1983)
\end{thebibliography}
\end{document}